\documentstyle[preprint,prl,aps]{revtex}
\tighten
\begin{document}

\draft

\title{The Origin of the Wigner Energy}

\author{
W. Satu{\l}a$^{1-3}$,
D.J. Dean$^{4}$,
J. Gary$^{2}$\thanks{Science Alliance summer student
from University of Missouri - Rolla.},
S. Mizutori$^{1,4}$,
W. Nazarewicz$^{2-4}$
}

 \address{
 $^1$Joint Institute for Heavy Ion Research,
    Oak Ridge National Laboratory,
    P.O. Box 2008, \\ Oak Ridge,   TN 37831, U.S.A. \\
 $^2$Department of Physics, University of Tennessee,
				      Knoxville, TN 37996, U.S.A.\\
 $^3$Institute of Theoretical Physics, Warsaw University,
  ul.   Ho\.za 69,  PL-00681, Warsaw, Poland         \\
 $^4$Physics Division, Oak Ridge National Laboratory,
    P.O. Box 2008, Oak Ridge,   TN 37831, U.S.A. 
 }

\maketitle
\begin{abstract}
Surfaces of experimental masses 
of even-even and odd-odd nuclei
exhibit a sharp slope discontinuity
at  $N$=$Z$. This cusp (Wigner energy), reflecting an
additional binding in  nuclei with neutrons and protons
occupying the same shell model orbitals, is usually
attributed to  neutron-proton pairing
correlations.
A method is developed
to extract the Wigner term from experimental
data. Both
empirical arguments and 
shell-model calculations
suggest that
the Wigner term  can be traced back to the isospin
$T$=0  part of nuclear interaction.
Our calculations
reveal the rather complex mechanism
responsible for the 
nuclear binding around the $N$=$Z$ line.
In particular,
we find that the Wigner term cannot be
solely explained in terms of correlations 
between  the neutron-proton
  $J$=1, $T$=0  (deuteron-like) pairs.
\end{abstract}


\narrowtext

One of the most interesting avenues in nuclear structure
research is the study of nuclear properties   near the particle
drip lines. Theoretically, due to their unusual shell structure
and weak binding, nuclei with highest and lowest
 $N/Z$ ratios represent a
unique challenge for the many-body problem.  

This Letter concerns the proton-rich border of nuclear binding
($N/Z$$\approx$1). 
The proton drip line is relatively well known experimentally,
and the discovery of $^{100}$Sn indicates
the potential that radioactive nuclear beams will access all
bound $N$=$Z$ nuclei and their bound and unbound neighbors. This
opens up the possibility of nuclear structure studies of
medium-mass and heavy systems with $N$$\approx$$Z$.

There are many aspects of nuclear structure which make physics
near the $N$=$Z$ line very interesting: proton and diproton
emission, isospin mixing, superallowed beta decay, and nuclear
constraints for the astrophysical $rp$-process - all these are
timely experimental and theoretical themes. A unique aspect of
nuclei with  $N$=$Z$ is that neutrons and protons occupy the same
shell-model orbitals. Consequently, 
the large spatial overlaps between neutron
and proton single-particle wave functions are expected to
enhance neutron-proton ($np$) correlations,  especially the $np$
pairing.

At present, it is not clear what the specific experimental
fingerprints of the $np$ pairing are, whether the 
$np$ correlations
are strong enough to form
 a static condensate, and what their 
main building blocks are. 
Most of our knowledge about nuclear pairing 
 comes from nuclei with a  sizable neutron excess where
the isospin $T$=1 neutron-neutron ($nn$) and proton-proton 
($pp$) pairing  dominate. Now, for the first time,
there is an experimental opportunity to explore
nuclear  systems in the
vicinity of the $N$=$Z$ line which have
 {\em many} valence $np$ pairs;
that is,  to probe the interplay between the like-particle and
neutron-proton ($T$=0,1, $T_z$=0) pairing channels.

This novel situation calls for  the generalization of
established theoretical models  of nuclear pairing.  In spite of
several early attempts  to extend  the
independent quasi-particle formalism to incorporate the effect
of $np$ correlations in light nuclei
\cite{Gos64,Gos65,Che67,Wol71} (see Ref.~\cite{Goo79} for 
an early 
review), no symmetry-unrestricted calculations  for $np$
pairing, based on the quasi-particle theory, 
have been carried out.

So far, the strongest evidence for $np$ pairing comes from the
masses  of $N$=$Z$ nuclei. An additional binding (the so-called
Wigner energy) found in these nuclei  manifests itself as a
spike in the isobaric mass parabola as a function of 
$T_z$=$\frac{1}{2}(N-Z)$ (see
the review \cite{Zel96} and Refs.  quoted therein).  Recently,
it has been  demonstrated that the $T$=0 $np$-pair correlations,
treated within the framework of the generalized quasi-particle
theory,  may produce such a cusp at $N$=$Z$ \cite{Sat97}. 
Gross estimates of the  magnitude of the Wigner energy
 come from a large-scale fit to experimental binding energies
with the 
 macroscopic-microscopic
approach \cite{Mye66,Kra79} and  from the analysis
of experimental masses \cite{Jen84}.
  In this study, we present a
technique to extract the Wigner energy directly from the
experimental data, and  give empirical arguments that this
energy originates primarily from the $T$=0  part of the
effective interaction. To obtain deeper insight into the
structure of the Wigner term, we apply the 
nuclear shell model to nuclei from the $sd$ and $fp$ shells.

In the semi-empirical mass formulae 
(see e.g. Ref.~\cite{Kra79}), an additional
binding due to the $np$-pair correlations
is usually parameterized as
\begin{equation}\label{eq1}
B^{\rm pair}_{np}= -\epsilon_{np}(A)\pi_{np}+E_W,
\end{equation}
where $\pi_{np}=\frac{1}{4}(1-\pi_n)(1-\pi_p)$
and  $\pi_n=(-1)^N$ and $\pi_p=(-1)^Z$ being
the {nucleon-number} parities.
The first contribution to the $np$ pairing energy
 in Eq.~(\ref{eq1}), $\epsilon_{np}$,
represents
an additional  binding
due to the residual interaction between
the two odd nucleons in an odd-odd nucleus. 

The second contribution to $B^{\rm pair}_{np}$,
$E_W$,   dubbed
the Wigner energy, is believed to
represent the energy of collective
$np$-pairing correlations. 
It can be decomposed into two parts:
\begin{equation}\label{EW1}
E_W = W(A)|N-Z| + d(A)\pi_{np}\delta_{NZ},
\end{equation}
The $|N-Z|$-dependence in Eq.~(\ref{EW1})
was first introduced by Wigner~\cite{Wig37} in
his analysis of the SU(4) spin-isospin symmetry of 
nuclear forces. In the supermultiplet approximation, there
appears
a term in the nuclear mass formula which is proportional to
$T_{\rm gs}(T_{\rm gs}+4)$, 
where $T_{\rm gs}$ denotes the isospin
of the ground state.
Empirically, $T_{\rm gs}$=$|T_z|$ for most nuclei except for
heavy odd-odd  $N$=$Z$ systems \cite{Jan65,Zel76}. 
Although the experimental
data indicate that the 
SU(4) symmetry is severely broken, 
and the masses behave according to the  
$T_{\rm gs}(T_{\rm gs}+1)$
dependence \cite{Jan65,Jen84}, 
the expression of Eq.~(\ref{EW1}) for the Wigner energy
is still very useful.
In particular, it accounts for a non-analytic behavior
of nuclear masses when an isobaric chain crosses the $N$=$Z$ 
line. An additional contribution to the Wigner term,
the $d$-term in  Eq.~(\ref{EW1}),
represents a 
correction for $N$=$Z$ odd-odd nuclei.
Theoretical justification of Eq.~(\ref{EW1}) has been given
in terms of basic properties of effective
shell-model interactions \cite{Zel96,Tal62}, and also by
using simple arguments based on the number of valence  
$np$-pairs \cite{Jen84,Mye77}. The estimates based on the 
number of $np$ pairs in identical spatial orbits suggest that
the ratio $d/W$ is constant and equal to one \cite{Mye77}.
A  different estimate has been given in 
 Ref.~\cite{Jen84}:
$d/W$=0.56$\pm$0.27.

In order to extract empirical information on the magnitude of
pairing correlations, one usually applies mass relations
(indicators, filters) defined as specific combinations of
nuclear binding energies of neighboring nuclei \cite{Jen84}. 
The use of these filters is based on the following assumptions:
({\it i\/}) the nuclear binding
energy  may be decomposed into a part $\tilde{B}$ 
which varies smoothly with $N$ and $Z$ (at
least away from the closed shells)
and a fluctuating (and possibly non-analytic) term;
and ({\it ii\/}) the pairing
energy may be parameterized as a smooth function of $N$ and $Z$
to reproduce global (large-scale) behavior. 
The important feature of mass indicators, based on
finite-differences of the nuclear binding
energy with respect to $N$ and $Z$, is that they remove
the smooth  energy background $\tilde{B}$ up to
third \cite{Jen84,Zel65} (or higher \cite{Mol92}) orders, and
 isolate  specific parts of the pairing energy.
However,  the interpretation of 
results depends on  the particular indicator used
and  the assumed 
parameterization of pairing energy.

The magnitude of the  Wigner term  is non-negligible only
in the closest vicinity of the $N$=$Z$ line. However, since
pairing indicators  are given by finite differences,
perturbations  at  $N$=$Z$ might disturb
the physical interpretation of extracted quantities.  For
example, the well-known four-point odd-even mass difference
formulae for the   $nn$ and $pp$ pairing gaps are quite strongly
perturbed near the  $N$=$Z$ line \cite{Sat97a}.  Consequently,
to isolate interesting effects, other indicators,
involving more nuclei, must be used.

 An  indicator particularly useful for investigating the
$np$ correlation  energy is the double-difference
formula  of Ref.~\cite{Zha89}:
 \begin{eqnarray}\label{vnp}
 \delta V_{np} (N,Z) & = &
{1 \over 4} \left\{ B(N,Z) - B(N-2,Z) -B(N,Z-2)
 + B(N-2,Z-2)\right\}\nonumber \\
 & \approx & {\partial^2B\over{\partial N\partial Z}}.
 \end{eqnarray}
This indicator  belongs to a different class than those analyzed
in Ref.~\cite{Jen84}. First, it involves binding energies of
nuclei having the same nucleon-number parity
and, therefore, it is  sensitive to the
Wigner energy \cite{Bre90}.
Second, indicator (\ref{vnp})  does not remove
the second-order contribution to the smooth energy
proportional to $NZ$.
As a consequence, for nuclei with $T_z>1$, $\delta V_{np}$ 
mainly probes the
symmetry energy term, $\sim (N-Z)^2/A$, and for nuclei 
with $T_z\le 1$ it contains {\em both} Wigner energy and 
symmetry energy contributions.
The nuclear symmetry energy 
is well described by a self-consistent
mean-field theory 
 based on the realistic effective 
particle-hole interaction.
Indeed, the application of indicator (\ref{vnp}) to the
theoretical self-consistent
mass tables
 using either a spherical 
Hartree-Fock-Bogoliubov model (HFB) \cite{Jacek}
or a deformed
extended Thomas-Fermi plus Strutinsky-integral model (ETFSI)
\cite{ETFSI}  gives results very
consistent with the experimental values of  $\delta V_{np}$
for nuclei with $T_z>1$ \cite{Sat97a}.
 This simple test
indicates that the symmetry energy is rather insensitive to
shell effects. Its strength  
 smoothly varies with $A$ and is practically
 $T_z$-independent. Since, in this study,
we are mainly interested in the Wigner 
energy, another indicator has to be employed that 
allows for
filtering out the symmetry energy contribution. 
This can be achieved by considering combinations of
$ \delta V_{np}$ values, as discussed in the following.

In this work,
the Wigner energy coefficient $W$ in an even-even nucleus
$Z$=$N$=$\frac{A}{2}$ has been  extracted by means of 
the indicator:
\begin{equation}\label{filtrw}
W(A) =   \delta V_{np}\left(\frac{A}{2},\frac{A}{2}\right)  -
{1\over 2} \left[ \delta V_{np}\left(\frac{A}{2},\frac{A}{2}-2
\right) +
\delta V_{np}\left(\frac{A}{2}+2,\frac{A}{2}\right) \right].
\end{equation}
The $d$-term  in
 an odd-odd  nucleus,
$Z$=$N$=$\frac{A}{2}$  [Eq.~(\ref{EW1})]
can be extracted using
another indicator:
\begin{equation}\label{filtrd}
d(A) =  
2\left[ \delta V_{np}\left(\frac{A}{2},\frac{A}{2}-2\right) +
\delta V_{np}\left(\frac{A}{2}+2,\frac{A}{2}\right) \right] -
4\delta V_{np}\left(\frac{A}{2}+1,\frac{A}{2}-1\right). 
\end{equation}
Although the recipe for
these third-order mass difference indicators
is not unique, the results appear to be very weakly dependent 
on the particular prescription used \cite{Sat97a}.

The experimental values of both contributions to the Wigner
energy are shown in Fig.~\ref{fig1}. The values of $W$
 decrease rather smoothly with $A$, showing
characteristic oscillations that are related to the
shell effects. Indeed, the filling of $1d_{5/2}2s_{1/2}$ shells
results in a decrease of $W$ with the local
minimum near $A$$\sim$34. Subsequent filling of the $1d_{3/2}$
shell tends to increase $W$. The next minimum in $W$ appears
at the middle of the $1f_{7/2}$ shell.
Table~\ref{t1} shows the results of a least-squares fit to 
$W$ assuming $W(A)=a_W/A^\alpha$. The actual 
minimum corresponds to $\alpha$$\approx$0.95; 
it is very close to 
the standard value of $\alpha$=1. The extracted
strength $a_W$$\approx$47\,MeV 
is considerably larger than the one used in the
semi-empirical mass formula of Ref.~\cite{Kra79} 
($a_W$=30\,MeV), but is  consistent with the value of
Ref.~\cite{Jen84}
($a_W$=43$\pm5$\,MeV).

As seen in  Fig.~\ref{fig1}, for the $sd$ nuclei,
the behavior of the $d$-term follows 
rather closely that of $W$, but
this nice agreement is lost for the heavier systems.
However,
it is  known experimentally (cf. discussion
in Ref.~\cite{Zel96} and references quoted
therein) that as a rule the nuclear ground states have isospin
$T_{\rm gs}$=$|N-Z|/2$, except for heavy $N$=$Z$ 
odd-odd nuclei. Indeed,
$^{34}$Cl and nuclei heavier than $^{40}$Ca 
(except for $^{58}$Cu)
have ground states of isospin $T$=1. 
Replacing, in our analysis, the
ground-state binding 
energies of these odd-odd $N$=$Z$, $T$=1 nuclei by
the binding energies of their lowest $T$=0 states
 leads to modified $d$-values
marked by triangles in Fig.~\ref{fig1}. The resulting
values 
follow very closely the values of $W$; i.e.
 $d_{T=0}(A)/W(A)\approx 1$, in a nice
agreement with the simple arguments of Ref.~\cite{Mye77}. This
observation provides an  important experimental argument
that both terms constituting $E_W$  have the same microscopic
origin. 
In particular, assuming that the structure of the lowest $T$=0
{\em states} 
in odd-odd $N$=$Z$ nuclei is strongly influenced by the
effective interaction between $T$=0 {\em pairs}, the close
similarity between  $d_{T=0}$ and $W(A)$  indicates that 
the $T$=0 component
of the nuclear interaction  is responsible
for the Wigner term.

The role of the $T$=0 part of the interaction
on the presence of binding-energy
irregularities in $\delta V_{np}$
near the  $N$=$Z$ has been recognized
in Ref.~\cite{Bre90}. They nicely
illuminated this point by demonstrating that in
the 
shell-model calculations 
for $sd$ nuclei using the
surface delta interaction (SDI),   it is 
indeed the $T$=0
part  of SDI that give rise to
the cusp in $\delta V_{np}$.

In order to investigate the role 
played by the $T$=0 component
for the Wigner energy, and to study the relative 
importance  of various 
$np$ pairs,
we performed shell-model
calculations in both the $1f2p$ and $1d2s$ shell-model spaces using
the shell-model code ANTOINE \cite{antoine}.
In the $fp$-shell we use the KB3 effective interaction of
Ref.~\cite{Pov81}, while for the $sd$-shell nuclei
we use the USD effective interaction of Ref.~\cite{Wil84}.
The results of the calculations are displayed in Figs.~\ref{fig2}
and \ref{fig3}.

The insert in Fig.~\ref{fig2}
shows the calculated binding energies (solid circles)
for the $A$=48 isobaric chain of even-even $fp$ nuclei
relative to the $N$=$Z$ nucleus $^{48}$Cr.
The Wigner term  gives rise  to  an
additional binding  for the $N$=$Z$ nuclei.
In Fig.~\ref{fig2}, this  is
  represented as a deviation, $\varepsilon_W$,
from a smooth parabola 
fitted
to the $N$$\ne$$Z$ isobars.
Note, however, that the quantity,
$\varepsilon_W$, is not a realistic estimate
of the Wigner energy $E_W$, but a rather schematic
measure of the cusp\footnote{In
reality, there is some contribution from the collective
$np$-pairing effects  also in nuclei with 
 $|N-Z|$$\geq$4,
and, due to the isospin invariance of the interaction, 
the isotopic chains of even-even nuclei considered in
 Fig.~\ref{fig2} contain only
three independent  masses. Hence,
the simple parabolic fit does not
represent the realistic  symmetry-energy dependence.}.

To visualize the influence of
the $T$=0 part of the effective nuclear
interaction on the Wigner term,
we have performed a set of shell-model
calculations while {\em switching off}
 sequentially the $J=1,
2,..., J_{\rm max}$,
$T$=0 two-body matrix elements $\langle 
j_1j_2JT|\hat{H}|j_1'j_2'JT\rangle$
of the shell-model Hamiltonian $\hat{H}$
for different values of $J_{\rm max}$.
Figure~\ref{fig2} shows a  ratio
$\varepsilon_W/\varepsilon_W^{\rm total}$,
where $\varepsilon_W^{\rm total}$ denotes the 
result of full shell-model
calculations versus  $J_{\rm max}$. The calculations were
performed for 
two representative examples, namely, the $fp$-shell
nucleus $^{48}$Cr and the $sd$-shell nucleus $^{24}$Mg.
The  largest contribution to the Wigner energy comes from
the part of the $T$=0 interaction between 
deuteron-like ($J$=1) and
`stretched' [$J$=5 ($sd$) and 7 ($pf$)] pairs.
The importance of these matrix elements
is well known; it is precisely for  $J$=1 and 
stretched  pair-states
that experimentally determined effective 
$np$  $T$=0 
interactions are strongest \cite{Ana71,Sch71,Mol75}. 
Note also that
the deuteron-like correlations contribute more strongly
to $\varepsilon_W$
in $sd$-nuclei than in $fp$-nuclei, and that matrix
elements corresponding to  intermediate values of $J$
give non-negligible contributions.
This reveals the complex structure of the Wigner energy, and
suggests that models which ignore high-$J$ components of
the $np$ interaction (e.g., by considering only
$J$=0, $T$=1 and $J$=1, $T$=0 $np$ pairs \cite{Eva81})
are not too useful for discussing the
actual  $np$ pair
correlations.

Since  in  the
self-consistent models based on 
the realistic effective interactions
the explicit $np$ coupling is ignored, they cannot reproduce 
the  Wigner term
(see, however, Ref. \cite{Sat97};
for the recent extension of the HFB method to $np$ pairing, see 
Ref. \cite{Per96}).
 The  successful ETFSI mass formula,
based on Skyrme-like particle-hole interaction and
monopole-pairing interaction between like nucleons,
 accounts well for 
most effects attributed to the particle-hole channel, e.g., 
shell 
effects and  deformations. However, all those are filtered out
by indicator 
(\ref{filtrw}). This is illustrated in the insert to
Fig.~\ref{fig3}
which shows that $W$$\approx$0 in ETFSI.

The $0\hbar\omega$ shell model, on the other hand, 
 reproduces well experimental values of $W$
 both in the $sd$- as well as in the $fp$-shell
(circles 
 in Fig.~\ref{fig3}).  The 
results of calculations
 with the
($J$=1, $T$=0) two-body matrix elements removed,
 marked by
triangles in Fig.~\ref{fig3},
demonstrate that the impact of the 
deuteron-like correlations on the strength of the Wigner term
is stronger in
$sd$-nuclei than in $fp$-nuclei (cf. also Fig.~\ref{fig2}).
We also note the strong
variations 
in the region of  $A$$\sim$32 where
$W$ becomes locally negative. 
These variations
are associated with the subsequent filling of the upper part of
the $1d_{5/2}$, the $2s_{1/2}$, and of the bottom part of the
$1d_{3/2}$ shells. In the extreme shell-model picture, 
both  $1d_{5/2}$ and  $2s_{1/2}$ shells are occupied for
$Z$=$N$=16, and the  $d_{3/2}$ shell
is well separated. Consequently,
 the important contribution to $W$ comes from  
$T$=0 matrix elements with $j_n$$\ne$$j_p$). 

Removing all $T$=0 matrix elements with
$J$=1,3,5,7 ($J_{\rm max}^{\rm odd}$=7 variant), i.e. those
which are predominantly due to the coupling 
between neutrons and protons
{\em in identical orbits},
 washes out $W$ almost
completely. An exception is the upper part of the $sd$-shell
($d_{3/2}$ subshell) where the value of $W$
is reduced by about
50-60\%.  
Intuitively, below $^{40}$Ca
one  can expect a quenching of
phase-space available for scattering the  
$np$-pairs in identical orbits.  Simultaneously, one would expect 
a slight increase of
the role of  $j_n$$\ne$$j_p$  $np$-pairs. Indeed, removing the
remaining $T$=0 matrix elements 
with even values of $J$
($J_{\rm max}$=7 variant) has the
largest impact on the heaviest 
$sd$-shell nuclei.

We briefly comment on the first term, $\varepsilon_{np}(A)$, in
Eq.~(\ref{eq1}).
 The magnitude of the residual
$np$ correlation between the last nucleons
in an odd-odd system
  may be extracted from
experimental masses using the
pairing indicators $\Delta_{np(pn)}$
proposed in Ref.~\cite{Jen84}. In the semi-empirical mass
formulae,  the most common parameterization
of $\varepsilon_{np}(A)$
 assumes a volume-like
dependence of $\varepsilon_{np}(A)\approx \varepsilon / A$ with
$\varepsilon$$\approx$20-40\,MeV \cite{Kra79,Jen84}.  Our analysis
performed for $A$$\geq$20
 shows that the $\varepsilon_{np}(A)$ is almost
mass-independent  for nuclei around the $N$=$Z$ line (with
$|N-Z|\leq 3$), and is of the order of
 $\varepsilon_{np}(A)$$\approx$500\,keV.
 Therefore, particularly for light nuclei,
 the magnitude of  $\varepsilon_{np}$
 is 2-3 times smaller than suggested
in Refs.~\cite{Kra79,Jen84}.
For heavier nuclei with larger
neutron excess,  $\varepsilon_{np}$
 is systematically weaker and approaches the value of
 $\sim$300\,keV.

In summary, we performed a systematic study of $np$-pair
correlations in $N$$\sim$$Z$ nuclei. Our main emphasis has been
the analysis of the  Wigner term in terms of 
$np$ pairs. 
 The experimental data are consistent with
a  two-term structure given by Eq.~(\ref{EW1}). We developed a
reliable method which allows for the  extraction of both $W$ and
$d$ components of the Wigner energy
 from experimental data.  Experimental values
are consistent with the simple relation $d_{T=0}/W \approx 1$. 
This is an  empirical argument that the  Wigner term
emerges  from the $T$=0 part of the
$np$ pair interaction. The
presence of an extra correction to the binding energy of odd-odd
nuclei (the $d$-term) may have  important consequences for
the  independent quasi-particle theory of $np$ pairing where
even-even and odd-odd nuclei are treated on the same footing.

To investigate the  microscopic structure of the Wigner term,
we performed shell-model calculations for
even-even nuclei while attenuating, in various ways, the
$T$=0
part of the effective 
shell-model interaction.  This study confirms that
the bulk part of Wigner energy 
comes from the $T$=0 $np$ pairs.
We also found that  the contribution from
deuteron-like ($J$=1, $T$=0) pairs 
is by no means dominant;
it is similar to,  and near the middle of the $f_{7/2}$-shell
even weaker than, the contribution from maximally aligned pairs.
The contributions from the $T$=0 pairs with intermediate
spins are also non-negligible, especially
in the
upper part of the $sd$-shell.
Finally, 
 we found that the residual
interaction between the valence proton and the valence neutron in
an odd-odd nucleus  exhibits
a very weak mass dependence.

\acknowledgments

The authors would like to thank Jacek Dobaczewski for useful 
comments.
This research was supported in part by the U.S. Department of
Energy under Contract Nos. 
DE-FG02-96ER40963
(University
of Tennessee), DE-FG05-87ER40361 (Joint Institute for Heavy
Ion Research), DE-AC05-96OR22464 with Lockheed Martin Energy
Research Corp. (Oak Ridge National Laboratory),  
and  by the Polish Committee for Scientific
Research under Contract No.~2~P03B~034~08.  DJD
acknowledges a Wigner fellowship from ORNL.

\begin{table}
\caption[A]{
Results of the least-squares fits for the  Wigner energy
strength
$W(A) = a_W/A^\alpha$  for
$\alpha$=1/2, 2/3, and 1. 
 The actual minimum of the mean standard
deviation $\sigma$ 
corresponds to $\alpha$=0.92 ($\sigma$$\approx$0.196). 
All data points
  with $Z$$\ge$10
shown in Fig.~\protect\ref{fig1} were considered in the fit.
}
\begin{tabular}{rrr}
 $\alpha$  &   $a_W$  &  $\sigma_W$   \\
  & MeV & MeV   \\
\tableline
1/2  &  8  &   0.239     \\
2/3  & 14  &   0.213     \\
  1  & 47  &   0.196   
\end{tabular}
\label{t1}
\end{table}

\begin{figure}[t]
\caption{
Experimental values of $W$ (filled circles) and $d$
 [Eq.~(\protect\ref{filtrd}), open circles]
in $N$=$Z$ nuclei
extracted from measured binding energies 
\protect\cite{Aud95}. For even-even 
nuclei, the values
of $W$ were obtained from 
Eq.~(\protect\ref{filtrw}), while the values
for   odd-odd
nuclei were obtained by means of a
 similar indicator, involving
binding energies of $N$=$Z$ and $N$=$Z$+2 even-even cores.  
The triangles mark the values of $d$  calculated
using experimental binding energies
of the lowest $T$=0 states in odd-odd nuclei. The solid line
represents the average value $W_{\rm av}$=47/$A$ (see text).
}
\label{fig1}
\end{figure}

\begin{figure}[t]
\caption{
Calculated
displacement $\varepsilon_W$ of the binding energy of 
$^{24}$Mg and $^{48}$Cr
from the average parabolic $(N-Z)^2$
behavior along an isobaric chain.
 Shell-model calculations were performed in the 
$0\hbar\omega$ configuration  space.
The results of 
calculations for the binding energies
of even-even nuclei
along  the $A$=48 chain (normalized to  $^{48}$Cr)
are shown in the insert.
The values of  $\varepsilon_W$ 
were obtained using the shell-model Hamiltonian
with
the $J=1,2,..,J_{\rm max}, T=0$ matrix elements removed.
For instance, the result for $J_{\rm  max}$=3
corresponds to the variant of calculations in which
all the two-body  matrix elements between states
$|j_1j_2 JT$=$0\rangle$ with $J$=1,2,3, were put to zero.   
The results are normalized to  
the full shell-model value 
$\varepsilon_W^{\rm total}$ ($J_{\rm  max}$=0). 
The Coulomb contribution to the binding energy has been 
disregarded.
}
\label{fig2}
\end{figure}

\begin{figure}[t]
\caption{
The strength of 
the Wigner term, $W$,
 extracted using binding energies calculated with the
$0\hbar\omega$
shell model.  Full shell-model calculations
(filled circles) 
 agree very well with experimental data (open circles).
The  results of shell-model
calculations with the
($J$=1, $T$=0) two-body matrix elements removed
($J_{\rm max}$=1 variant, triangles),
with all  $T$=0 matrix elements removed
($J_{\rm max}$=7 variant, diamonds),
and with ($J$=1,3,5,7, $T$=0) two-body matrix elements removed
($J_{\rm max}^{\rm odd}$=7 variant, squares)
are also shown. 
In the $fp$-shell, only two
points have been calculated due to practical limitations. Also a
bridge between $sd$ and $fp$ shells cannot be covered as it
requires interactions that include cross-shell matrix elements
which are currently  not  established.
The insert shows the  values of $W$ 
extracted from the ETFSI mass formula
\protect\cite{ETFSI}. They are practically zero for all nuclei
considered. 
}
\label{fig3}
\end{figure}

\end{document}